\documentclass[fleqn,10pt]{wlscirep}

\title{Magnetostatic interaction between two Bubble Skyrmions }

\author[1]{M. A. Castro}
\author[1]{D. Mancilla-Almonacid}
\author[2]{J. A. Valdivia}
\author[1,*]{S. Allende}

\affil[1]{Departamento de F\'\i sica, CEDENNA, Universidad de
Santiago de Chile, USACH, Av. Ecuador 3493, Santiago, Chile}

\affil[2]{Departamento de F\'\i sica, Facultad de Ciencias, Universidad de Chile, Casilla 653, 7800024, Santiago, Chile}

\affil[*]{sebastian.allende@usach.cl}

\begin{abstract}
A detailed analytic and numerical analysis of the interaction between two bubble skyrmions has been carried out. Results from micromagnetic calculations show a strong dependence of the parameters of the skyrmion magnetic profile as a function of the magnetostatic interaction. The magnetic core and edge-width sizes of the skyrmion increase or decrease depending on the relative position between the skyrmions and the uniaxial perpendicular anisotropy.  In particular, when a magnetic disk is over another, there is a transition from a Bloch-like skyrmion configuration to a N\'{e}el-like skyrmion configuration as the distance between the disks decreases, as a consequence of the  magnetostatic interaction. Therefore, it is possible to stabilize a bubble skyrmion with a N\'{e}el configuration without the Dzyaloshinskii-Moriya interaction.  Thus, these results can be used for the parameters control of the skyrmions in magnetic spintronic devices that need to use these configurations. 

\end{abstract}

\usepackage{amsfonts}
\usepackage{amsmath}
\usepackage{amssymb}
\usepackage{graphicx}
\usepackage{epstopdf}
\usepackage{todonotes}
\usepackage{color}
\usepackage{ulem}
\usepackage{tocloft} 
\usepackage[fleqn]{nccmath} 
\usepackage{lmodern}
\DeclareMathOperator{\sech}{sech}

\begin{document}

\flushbottom
\maketitle

\thispagestyle{empty}

\section*{Introduction}

  During the last decade, a great deal of attention has been focused on
the study of the magnetic skyrmions in magnetic structures because they
have potential applications in magnetic storage devices of high
density, spintronic devices, etc. \cite{Fert2013,Zhang2015, Iwasaki2013, SampaioJ.2013,Zhang2015b,Zhang2015c}. For example, in nanostructures such as nanodisks it is possible to find different type of skyrmions, like N\'{e}el, Bloch or bubble configurations, among others. The
N\'{e}el and Boch skyrmion configurations can be obtained by introducing a
Dzyaloshinskii-Moriya interaction due to the strong spin-orbit
coupling between two materials \cite{Fert2013,Zhang2015, SampaioJ.2013, Rohart2013}.  Similarly, the
  bubble skyrmions can be stabilized through an uniaxial magnetic
anisotropy perpendicular to the plane of the disk \cite{guslienko2015,Montoya2017,Montoya2017b,Wang2017,Castro2016}. 

It is interesting to note that the magnetic particles
  possess a long-range magnetostatic field, which is present in the
formation of a great variety of magnetic textures like vortices or
skyrmions.   Recently, arrays
of bubble skyrmions in nanodisks with perpendicular anisotropy have been
proposed for the implementation of spintronic devices \cite{Bttner2015,Mochizuki2017,Sun2013,Gilbert2015}. In these
systems, it should be emphasized that the interaction between the
skyrmions through the magnetostatic field can be strong depending on
their locations \cite{Zhang2015,Mller2017}. In terms of analysis, the interaction between bubble
skyrmions can be decomposed as the
magnetostatic field interaction of cores and edges. This
magnetostatic interaction could even influence their movements and may also
affect their magnetic structures \cite{Mller2017,Ding2015} affecting the operation of the device. Therefore it becomes necessary to study in detail the interaction between two bubble skyrmions.

Hence, in this paper, we study the magnetostatic interaction between two magnetic dots that have a magnetic bubble skyrmion. They are stabilized by an effective anisotropy without the Dzyaloshinskii-Moriya interaction. Specifically, we focus our attention on the skyrmion core and edge that vary in size as a function of the
  magnetostatic interaction between these two magnetic dots. Based on
  micromagnetic calculations and micromagnetic simulations, we have carried out numerical
  calculations, in which we have observed a strong variation of the
  parameters of the skyrmion magnetic profile. The magnetic
  core and the edge-width sizes of the skyrmion increase or decrease
  depending on the relative position between the skyrmions and the
  uniaxial perpendicular anisotropy. In particular, it is possible to stabilize bubble skyrmions with a N\'{e}el-like skyrmion magnetic profile when a magnetic disk is over another, in the absence of the Dzyaloshinskii-Moriya interaction. This transition from the Bloch-like skyrmion configuration to the N\'{e}el-like skyrmion configuration is due to the magnetostatic interaction between the magnetic disks. These results could be useful for
  the realization of future bubble skyrmion devices.

\section*{Theory}

\begin{figure}[h!]
\begin{center}
\includegraphics[width=12cm]{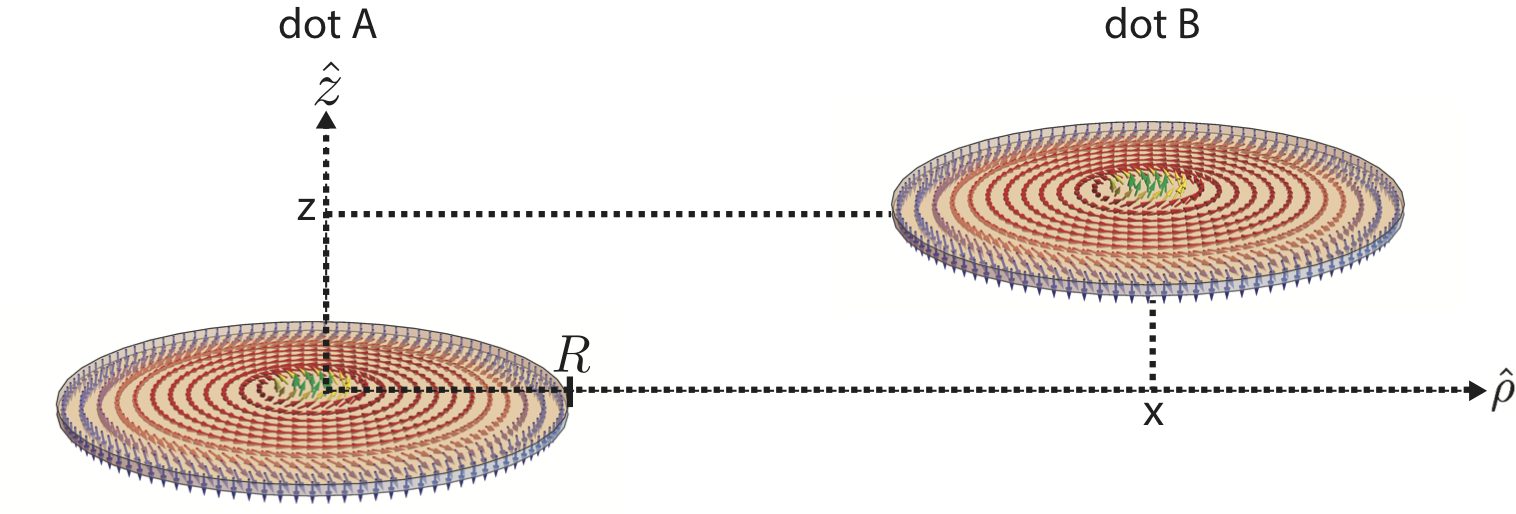}
\caption{Schematic representation of two bubble skyrmions separated, center to center, by a horizontal distance $x$ and a vertical distance $z$. They are coupled by the magnetostatic interaction.}
\label{fig1}
\end{center}
\end{figure}

We start with two dots that have a  magnetic Co/Pt bubble
  skyrmion configuration. These dots are separated, center to center, by
  a horizontal distance $x$ and a vertical distance $z$, as shown in
  Fig. \ref{fig1}.  The skyrmions are then allowed to
  interact through the magnetostatic interaction. Each magnetic dot
has a radius $R$, a height $H$, and an effective magnetic uniaxial anisotropy
perpendicular to the plane of the dot characterized by
$K_u>0$. The magnetic parameters for each
dot are $M_s=500$ kA/m and $A=1.5\times 10^{-11}$ J/m, so
  that the exchange length is equal to $L_{\text{ex}}=\sqrt{2A/\mu_0M_s^2}\approx 9.8$ nm \cite{CoPt}. We approach the study of these systems with the micromagnetic theory by using analytical and numerical calculations, and micromagnetic simulations. 
  
 The micromagnetic simulations are performed with the Object Oriented Micromagnetic Framework (OOMMF) code \cite{Oommf}.  We consider that each dot has a thickness $H=10$ nm and a radius $R=300$ nm,  a cubic mesh size of $2\times2\times2$ nm$^3$  and the Gilbert damping constant equal to $0.5$. To obtain the minimum energy configuration, we consider different initial states of the magnetization such as vortex, in plane, out of plane, and skyrmion configurations. To relax the system into the most stable configuration, we use the Euler method.

    \subsection{Horizontal separation between two magnetic disks with low anisotropy }
\begin{figure}[h!]
\begin{center}
\includegraphics[width=14cm]{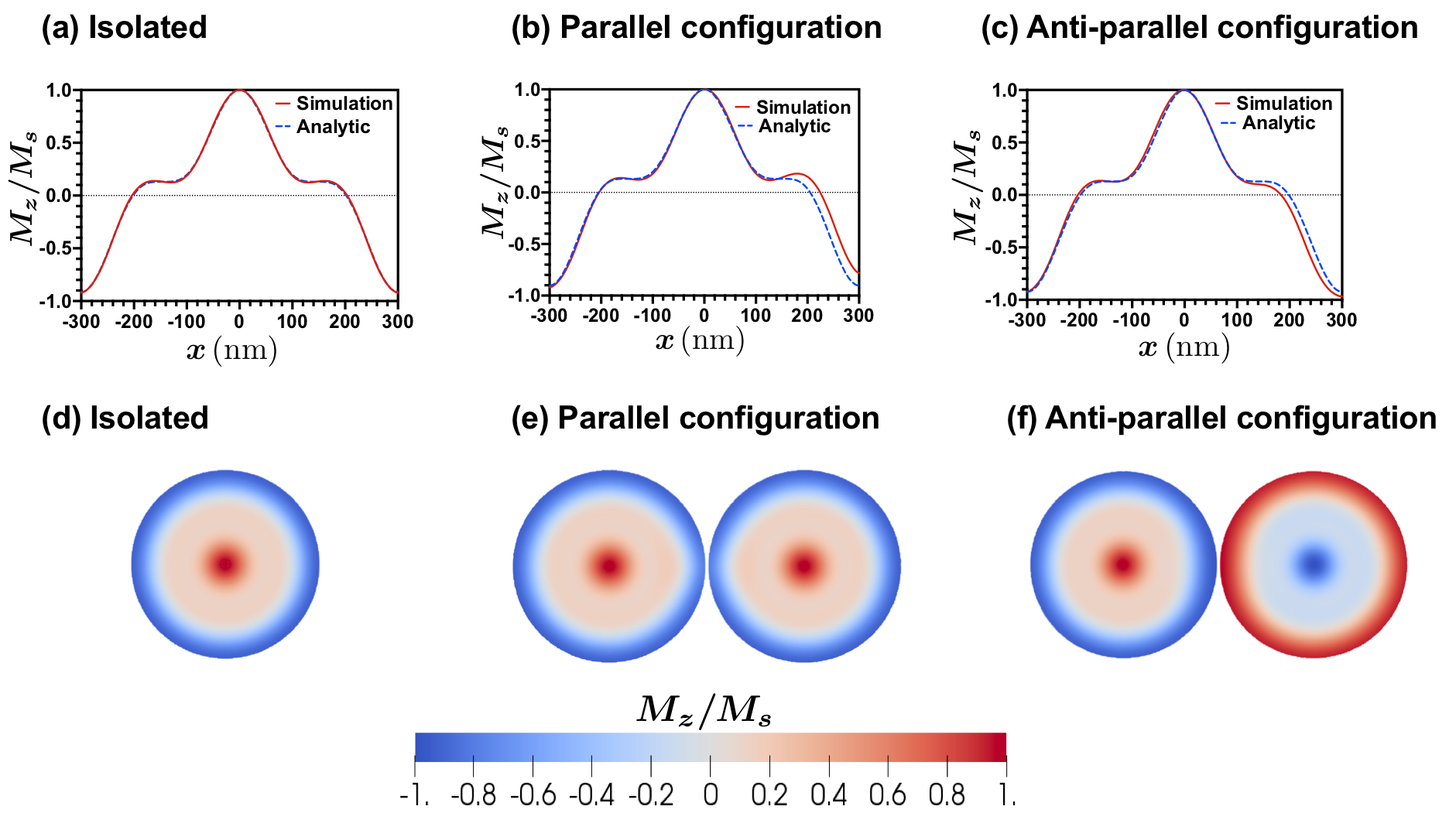}
\caption{ Comparison between the analytic magnetic profile given by Eq. (\ref{perfil}) and the magnetic profile obtained by the micromagnetic simulation with OOMMF with $R=300$ nm, $H=10$ nm, $K_u=143$ kJ/m$^3$, and $z=0$. Magnetic profile of (a) an isolated disk, (b) two parallel $B_1$ configuration with $x=610$ nm,  and (c) two antiparallel $B_1$ configuration with $x=610$ nm. In addition, Figures \ref{comparess}(d), \ref{comparess}(e), and \ref{comparess}(f) show a top view of the minimum energy configuration obtained by OOMMF for the  Figures \ref{comparess}(a), \ref{comparess}(b), and \ref{comparess}(c), respectively.}
\label{comparess}
\end{center}
\end{figure}

In the first place, we consider two disks,  with low magnetic anisotropy, separated by a horizontal distance  ($x>R$) and in the same plane ($z=0$). From the micromagnetic simulations, we propose a  magnetic profile of the form of a  Bloch-like skyrmion characterized by a magnetization that rotates in the plane and perpendicular to the radial direction, i.e., $\vec{M}\left(\vec{r}\right)=M_s[m_{\phi}\left(\rho\right)\hat{\phi}+m_{z}\left(\rho\right)\hat{z}]$,
where $M_s$ is the saturation magnetization of the dot, and $m_{\phi}^2(\rho)+m_{z}^2(\rho)=1$. 
Then,  the analytical and numerical calculations are done by parameterizing the magnetization function for the bubble skyrmion in cylindrical coordinates by
\cite{Novais2011,Castro2016}
\begin{ceqn}
\begin{equation}
m_{z}^{B_1}(\rho) = \left \{ \begin{matrix} \delta+\left(1-\delta \right) \left(1-\dfrac{\rho^{2}}{\alpha^{2}}\right)^{4} & 0<\rho \leq \alpha\\
 \delta & \alpha<\rho \leq \beta\\
\delta-g\left(1+\delta\right) \left(1-\dfrac{(R-\rho)^2}{(R-\beta)^2}\right)^4 &\beta<\rho \leq R
\end{matrix}\right.
\label{perfil} 
\end{equation}
\end{ceqn}
where $g$ is a parameter related to the maximum value of the $z$ component of the magnetization in the edge of the disk and takes the value between 0 and 1. $\delta$ is related to the plateau of the $z$ component of the magnetization observed in the OOMMF simulations. $\alpha$ and $\beta$ are related with the beginning and ending of the plateau, respectively [$m_{z}^{B_1}(\alpha)=m_{z}^{B_1}(\beta)=\delta$]. The abbreviation $B_1$ in the superindex of $m_{z}$ is used  with the aim of referring to a  Bloch-like skyrmion.
Figure \ref{comparess} illustrates the $z$-component of the magnetization of the magnetic profile for  $K_u=143$ kJ/m$^3$. The top row illustrates a comparison between the analytic magnetic profile given by Eq.~(\ref{perfil}) and the magnetic profile obtained by the micromagnetic simulation with OOMMF. The bottom row illustrates a top view of the magnetization obtained with OOMMF. Figures \ref{comparess}(a) and \ref{comparess}(d) considers an isolated magnetic dot. Figures \ref{comparess}(b) and \ref{comparess}(e) considers the strongest magnetostatic interaction between two disks with a parallel-configuration of the  magnetic bubbles with $x=610$ nm that we have studied. Figures \ref{comparess}(c) and \ref{comparess}(f) considers the strongest magnetostatic interaction between two disks with an anti parallel-configuration of the  magnetic bubbles with $x=610$ nm that we have studied.
The analytical magnetic profile shows a very good agreement with the micromagnetic simulations  when the disk is isolated.  We observe a difference between the analytical and numerical value where the $z$ component of the magnetization is equal to zero of approximately  6\%. Therefore, these OOMMF simulations suggest that the cylindrical angular variation  in the magnetic profile of these bubble skyrmions can be disregarded in first approach when they are interacting by the magnetostatic interaction, as suggest Figs. \ref{comparess}(d), \ref{comparess}(e), and \ref{comparess}(f). Hence, for simplicity, in the analytical analysis below we will consider that their magnetic profiles do not depend on the polar angle. Therefore, we use the magnetic profile given by Eq. (\ref{perfil}) when $z=0$ and $K_u=143$ kJ/m$^3$.

The total magnetic energy of two disks with the $B_1$ configuration, $E_{\text{tot}}^{B_1}$, is given by the sum
of the exchange, magnetostatic, and anisotropy energies, whose forms are suggested by the micromagnetic
theory\cite{aharoni}. The exchange energy for the bubble skyrmion with the $B_1$ configuration, $E_{\text{ex}}^{B_1}$, is given by \cite{Castro2016}
\begin{ceqn}
\begin{equation}
E_{\text{ex}}^{B_1}=2\pi HA\left[\int_{0}^{\alpha}f_1(\rho)\rho d\rho+\left(1-\delta ^2\right) \ln \left(\frac{\beta}{\alpha }\right)+\int_{\beta}^{R}f_2(\rho)\rho d\rho\right],
\label{eq2}
\end{equation}
\end{ceqn}
 where $A$ is the stiffness constant. The functions $f_1(\rho)$ and $f_2(\rho)$ in Eq.~(\ref{eq2}) are

    \begin{ceqn}
     \begin{equation}
   f_1(\rho)=\frac{1-\left[(1-\delta ) \zeta_1^4(\rho)+\delta \right]^2}{\rho ^2}+\frac{64 (1-\delta )^2 \rho ^2 \zeta_1^6(\rho)}{\alpha ^4 \left(1-\left[(1-\delta ) \zeta_1^4(\rho)+\delta \right]^2\right)},
   \end{equation}
   \end{ceqn}

      \begin{ceqn}
     \begin{equation}
   f_2(\rho)= \frac{1-\left[\delta-g (\delta+1) \zeta_2^4(\rho)\right]^2}{\rho ^2}+\frac{64 g^2 (\delta+1)^2 \left(1-\zeta_2(\rho)\right) \zeta_2^6(\rho)}{(R-\beta)^2 \left(1-\left[\delta-g (\delta+1) \zeta_2^4(\rho)\right]^2\right)},
   \end{equation}
   \end{ceqn}

   \noindent
   with $\zeta_1(\rho)=1-\rho^2/\alpha^2$ and  $\zeta_2(\rho)=1-(R-\rho)^2/(R-\beta)^2$.
   The magnetostatic contribution is given by the self-magnetostatic
   interaction for every dot defined by $E_{\text{m,self}}$, and the
   magnetostatic interaction between the dots called $E_{\text{m,int}}$. The
   self magnetostatic interaction is $E_{\text{m,self}}=(\mu_0/2)\int
   \vec{M}(\vec{r})\cdot \vec{\nabla}U_{\text{self}}(\vec{r})  dv$,
   where $U_{\text{self}}(\vec{r})$ is the magnetostatic potential in a dot
   due to the magnetization of the same magnetic dot\cite{aharoni}. Observing that there are not volumetric charges in the magnetic profile of
   Eq.~(\ref{perfil}) [$\vec{\nabla} \cdot \vec{M}(\vec{r})=0$],
   then $E_{\text{m,self}}$ has the form \cite{Castro2016}
     \begin{ceqn} 
   \begin{equation}
E_{\text{m,self}}^{B_1}=\pi\mu_{0}M_{s}^{2}\int_{0}^{\infty}dq\left[F_1\left(q\right)+F_2\left(q\right)\right]^2\left(1-e^{-qH}\right),
\label{skidip}
\end{equation}
   \end{ceqn}
   \noindent
   where $F_1\left(q\right)$ and $F_2\left(q\right)$ are
   \begin{ceqn}
    \begin{equation}
F_1\left(q\right)=F_{11}(q)J_1(q \beta)+F_{12}(q)J_1(q \alpha )+F_{13}(q)J_2(q \alpha ),
\end{equation}
\end{ceqn}
\begin{ceqn}
    \begin{equation}
F_2\left(q\right)=\int_{\beta}^{R} \rho J_{0}(q\rho)m_{z}^{B_1}(\rho)d\rho,
\label{nozero}
\end{equation}
\end{ceqn}
where $F_{11}(q)= \beta\delta/q$,  $F_{12}(q)= 384 (\delta -1)   \left(48-q^2 \alpha ^2\right)/(q^7 \alpha ^5)$, and $F_{13}(q)= 4608 (\delta -1)  \left(q^2 \alpha ^2-16\right)/(q^8 \alpha ^6)$.
The magnetostatic interaction between the two dots is given by
$E_{\text{m,int}}=\mu_0\int \vec{M}(\vec{r})\cdot
\vec{\nabla}U_{\text{int}}(\vec{r}) dv$, where $U_{\text{int}}(\vec{r})$ is
the magnetostatic potential in a dot due to the magnetization of the other magnetic
dot\cite{aharoni}. Then $E_{\text{m,int}}$ with the magnetic profile given by Eq. (\ref{perfil}) is equal to:
\begin{ceqn}
   \begin{equation}
E_{\text{m,int}}^{B_1}\left(x,z\right)=-\pi\mu_{0}M_{s}^{2}\sigma_A \sigma_B\int_{0}^{\infty}dq J_{0}(q x) \left[F_1\left(q\right)+F_2\left(q\right)\right]^2e^{-q(H+z)}g(q,H,z),
\label{skidip}
\end{equation}
\end{ceqn}
   
   \noindent
 where the letters $A$ and $B$ represent the dot $A$ and the dot
   $B$, respectively. $\sigma_A$ and $\sigma_B$ take values $\pm 1$, and their values
   define if the magnetization profile is given by Eq.~(\ref{perfil})
   (value $+1$) or minus the magnetization profile given by
   Eq.~(\ref{perfil}) (value $-1$). In essence we can parametrize the skyrmions with two orientations, namely, up or down. When $\sigma_A=\sigma_B$ the configuration is called parallel, and when $\sigma_A\neq \sigma_B$ the configuration is called anti-parallel. 
   The function $g(q,H,z)$ is:
\begin{ceqn}
   \begin{equation}
g(q,H,z) = \begin{cases}
\left(1-2 e^{q H}+e^{2q z}\right) &0\leq z<H\\
\left(e^{q H}-1\right)^2  &z\geq H
\end{cases}
\end{equation}
\end{ceqn} 
The anisotropy contribution, $E_{\text{ani}}$, is given by $E_{\text{ani}}=-K_u\int m_{z}^{2}(\rho) dv$.  Then, $E_{\text{ani}}$, with the magnetic profile given by Eq. (\ref{perfil}), is \cite{Castro2016}:
\begin{ceqn}
\begin{equation}
E_{\text{ani}}^{B_1}=-2\pi K_u H \left(\dfrac{\alpha^2}{18}-\dfrac{(R-\beta)^2g^2}{18}+\dfrac{32768}{109395}(R-\beta)g^2R+f_3(\delta)\right),
\label{skyani}
\end{equation} 
\end{ceqn} 
   \noindent
 where $f_3(\delta)$ is equal to
      \begin{align}
 &f_3(\delta)=f_{31}\delta +f_{32}\delta ^2 ,
  \end{align}
   
     \begin{align}
 &f_{31}= \left(-\frac{(R-\beta)^2 g^2}{9}+\frac{(R-\beta)^2 g}{5}+\frac{65536 (R-\beta) g^2 R}{109395}-\frac{256 (R-\beta) g R}{315}+\frac{4 \alpha ^2}{45}\right),
  \end{align}
      \begin{align}
 &f_{32}= \left(-\frac{(R-\beta)^2 g^2}{18}+\frac{(R-\beta)^2 g}{5}+\frac{32768 (R-\beta) g^2 R}{109395}-\frac{256 (R-\beta) g R}{315}+\frac{R^2}{2}-\frac{13 \alpha ^2}{90}\right).
  \end{align}
   
Hence, the expression of the total energy  of the system is equal to
\begin{ceqn}
\begin{equation}
E_{\text{tot}}^{B_1}=2E_{\text{ex}}^{B_1}+2E_{\text{m,self}}^{B_1}+2E_{\text{ani}}^{B_1}+E_{\text{m,int}}^{B_1}.
\label{skytot}
\end{equation} 
\end{ceqn}
This expression, Eq. (\ref{skytot}), depends on the parameters $x$, $z$, $\delta$, $g$, $\alpha$, and $\beta$.  Therefore, to obtain the energy
of the system, we need to minimize $E_{\text{tot}}^{B_1}$ as a function of the
parameters $\delta$, $g$, $\alpha$, and $\beta$; for a fixed $R$, $H$, $x$, $z$, and $K_u$.
 
  \noindent
  
    \subsection{Horizontal separation between two magnetic disks with high anisotropy }
  
\begin{figure}[h!]
\begin{center}
\includegraphics[width=14cm]{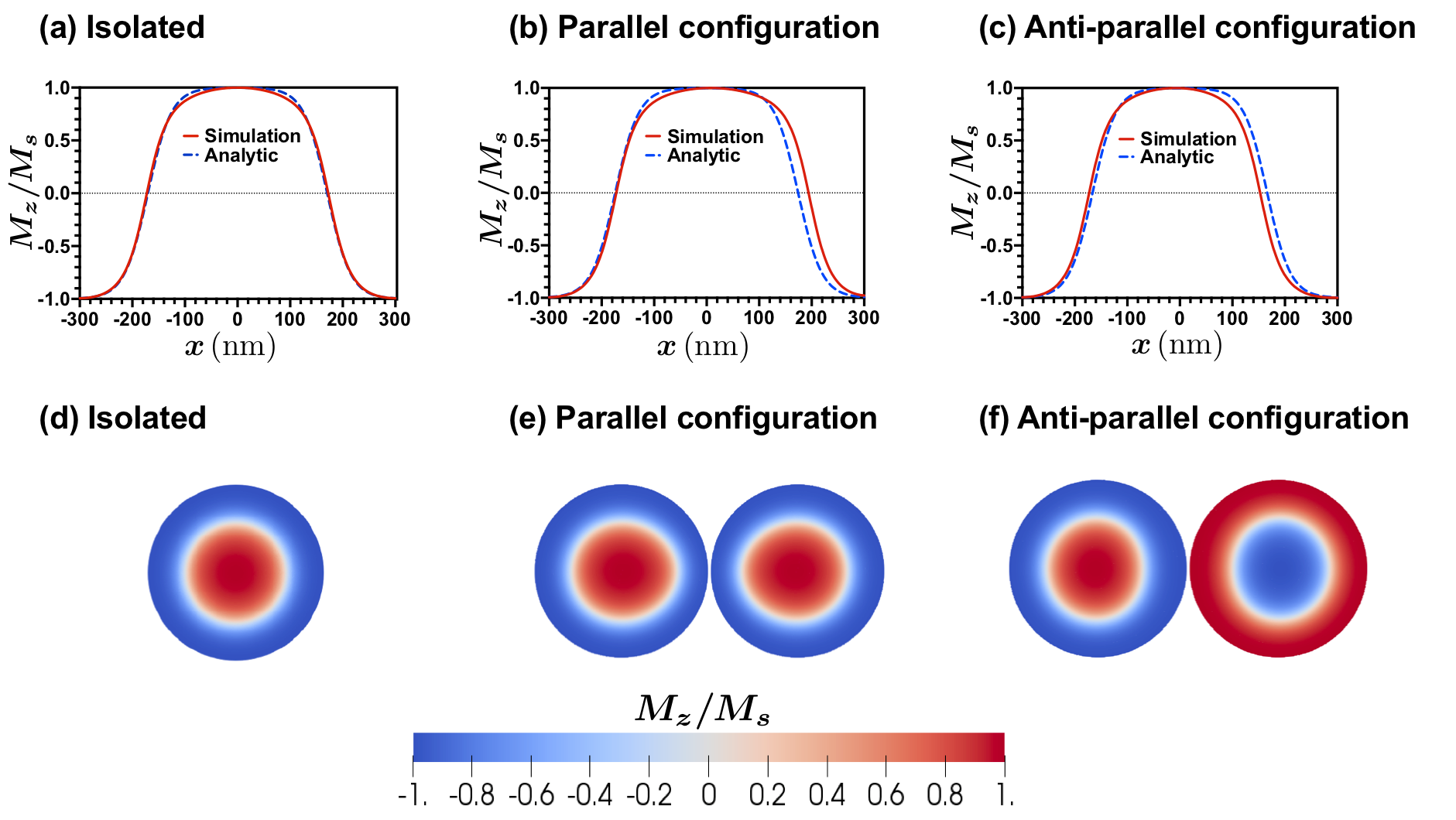}
\caption{ Comparison between the analytic magnetic profile given by Eq. (\ref{perfil2}) and the magnetic profile obtained by the micromagnetic simulation with OOMMF with $R=300$ nm, $H=10$ nm, $K_u=150$ kJ/m$^3$, and $z=0$. Magnetic profile of (a) an isolated disk, (b) two parallel $B_2$ configuration with $x=610$ nm,  and (c) two antiparallel $B_2$ configuration with $x=610$ nm. In addition, Figures \ref{comparea}(d), \ref{comparea}(e), and \ref{comparea}(f) show a top view of  the minimum energy configuration obtained by OOMMF for the  Figures \ref{comparea}(a), \ref{comparea}(b), and \ref{comparea}(c), respectively.}
\label{comparea}
\end{center}
\end{figure}

 In this section we consider two disks,  with high magnetic anisotropy, separated by a horizontal distance ($x>R$) and in the same plane ($z=0$). From the micromagnetic simulations, we propose a  magnetic profile of the form $\vec{M}\left(\vec{r}\right)=M_s[m_{\phi}\left(\rho\right)\hat{\phi}+m_{z}\left(\rho\right)\hat{z}]$, where the  magnetic profile for $m_{z}\left(\rho\right)$, is given by \cite{guslienko2015}  
  \begin{ceqn}
  \begin{equation}
  m_{z}^{B_2}(\rho) = \tanh\left(\frac{\rho-\gamma}{\Delta}\right),
  \label{perfil2}
  \end{equation}
  \end{ceqn}
  where $b=\gamma$ and $c=R-\gamma$ are the core and the edge-width of the magnetic bubble skyrmion, respectively.  The abbreviation $B_2$ in the superindex of $m_{z}$ is used  with the aim of referring to a  Bloch-like skyrmion configuration.
Figure \ref{comparea} illustrates the $z$-component of the magnetization of the magnetic profile for $R=300$ nm, $H=10$ nm, and $K_u=150$ kJ/m$^3$. The top row illustrates a comparison between the analytic magnetic profile given by Eq. (\ref{perfil2}) and the magnetic profile obtained by the micromagnetic simulation with OOMMF. The bottom row illustrates a top view of the magnetization obtained with OOMMF. Figures \ref{comparea}(a) and \ref{comparea}(d) considers an isolated magnetic bubble. Figures \ref{comparea}(b) and \ref{comparea}(e) consider the magnetostatic interaction between two disks with a parallel-configuration of the  magnetic bubbles with $x=610$ nm. Figures \ref{comparea}(c) and \ref{comparea}(f) consider the magnetostatic interaction between two disks with an anti parallel-configuration of the  magnetic bubbles with $x=610$ nm.
The analytical magnetic profile shows a very good agreement with the micromagnetic simulation when the disk is isolated.  We observe a difference between the analytical and numerical value where the $z$ component of the magnetization is equal to zero of approximate  7\%.
Therefore, these OOMMF simulations suggest that the cylindrical angular variation  in the magnetic profile of these bubble skyrmion can be disregarded when they are interacting by the magnetostatic interaction, as suggest  Figs. \ref{comparea}(d), \ref{comparea}(e), and \ref{comparea}(f). Hence, for simplicity, in the analytical analysis below we will consider that their magnetic profiles do not depend on the polar angle. Therefore, we use the magnetic profile given by Eq. (\ref{perfil2}) when $z=0$ and $K_u=150$ kJ/m$^3$.

  The total magnetic energy of the system with the $B_2$ configuration for the dots, $E_{\text{tot}}^{B_2}$, is given by the sum
of the exchange, magnetostatic, and anisotropy energies, whose form are suggested by the micromagnetic
theory\cite{aharoni}. The exchange energy for the bubble skyrmion configuration with the $B_2$ configuration, $E_{\text{ex}}^{B_2}$, is given by
\begin{ceqn}
\begin{equation}
E_{\text{ex}}^{B_2}=2\pi H A\int_{0}^{R}\frac{\left(\Delta^2+\rho^2\right)\sech^2\left(\frac{\gamma-\rho}{\Delta}\right) }{\Delta^2 \rho^2}\rho d\rho.
\label{eq2b}
\end{equation}
\end{ceqn}
   The self-magnetostatic energy with the $B_2$ configuration is
     \begin{ceqn} 
   \begin{equation}
E_{\text{m,self}}^{B_2}=\pi\mu_{0}M_{s}^{2}\int_{0}^{\infty}dq\left[\int_0^{R} J_0\left(q\rho\right)m_{z}^{B_2}(\rho)\rho d\rho\right]^2\left(1-e^{-qH}\right).
\label{skidipb}
\end{equation}
   \end{ceqn}
   \noindent
The magnetostatic interaction between the two dots is
\begin{ceqn}
   \begin{equation}
E_{\text{m,int}}^{B_2}\left(x,z\right)=-\pi\mu_{0}M_{s}^{2}\sigma_A \sigma_B\int_{0}^{\infty}dq J_{0}(q x) \left[\int_0^R J_0\left(q\rho\right)m_{z}^{B_2}(\rho)\rho d\rho\right]^2e^{-q(H+z)}g(q,H,z).
\label{skidipbb}
\end{equation}
\end{ceqn}  
   \noindent
The anisotropy contribution is:
\begin{ceqn}
\begin{equation}
E_{\text{ani}}^{B_2}=-\pi H R^2 K_u-2\pi H K_u \left(\Delta^2 \ln\left[\cosh\left(\frac{\gamma-R}{\Delta}\right) \sech\left(\frac{\gamma}{\Delta}\right)\right]+\Delta R \tanh\left(\frac{\gamma-R}{\Delta}\right)\right).
\label{skyanib}
\end{equation} 
\end{ceqn} 
   \noindent
Therefore, the total energy expression of the system with the $B_2$ configuration for the dots is equal to
\begin{ceqn}
\begin{equation}
E_{\text{tot}}^{B_2}=2E_{\text{ex}}^{B_2}+2E_{\text{m,self}}^{B_2}+2E_{\text{ani}}^{B_2}+E_{\text{m,int}}^{B_2}.
\label{skytotb}
\end{equation} 
\end{ceqn} 
This expression, Eq. (\ref{skytotb}), depends on the parameters $x$, $z$, $\gamma$, and $\Delta$.  Therefore, to obtain the energy
of the system, we need to minimize $E_{\text{tot}}^{B_2}$ as a function of the
parameters $\gamma$ and $\Delta$; for a fixed $R$, $H$, $x$, $z$, and $K_u$.

      \subsection{Vertical separation between two magnetic disks  }
\begin{figure}[h!]
\begin{center}
\includegraphics[width=14cm]{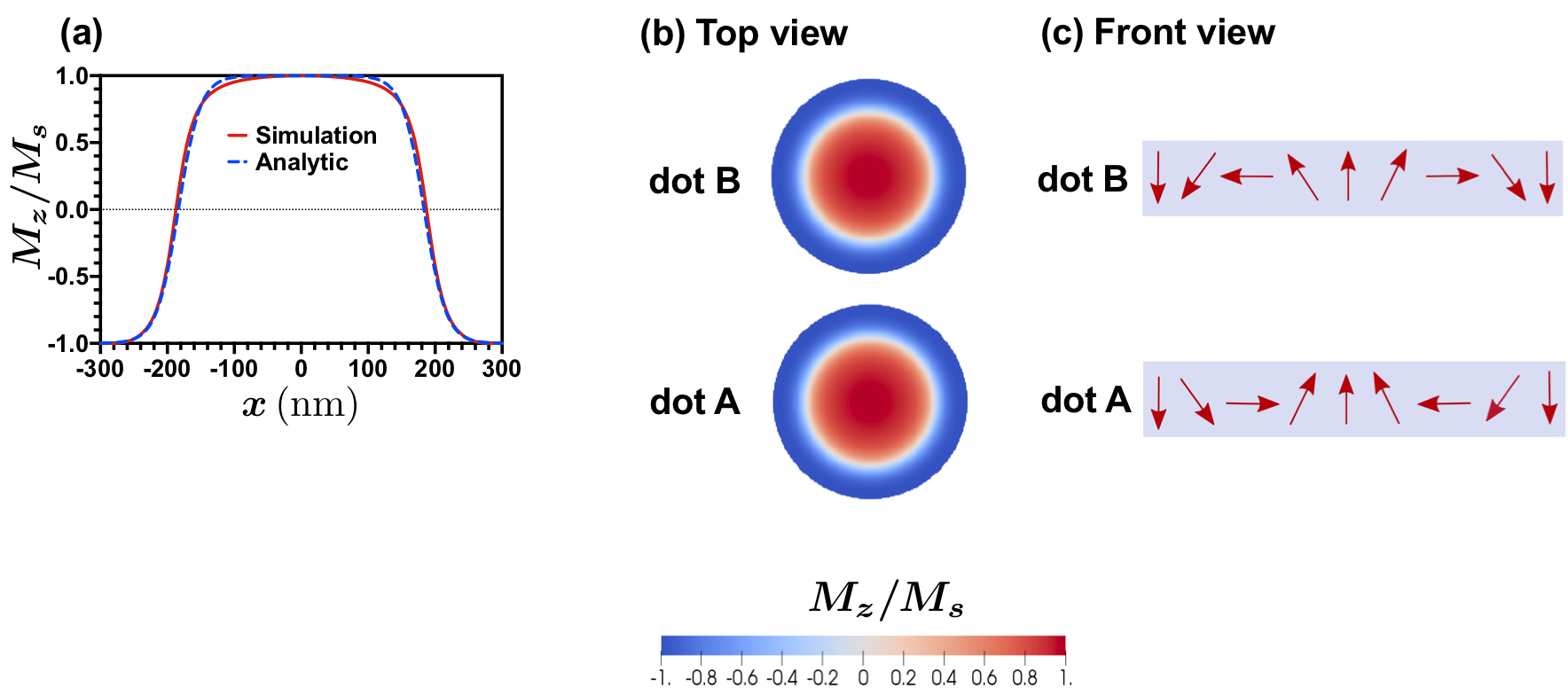}
\caption{ Magnetization of the minimum energy state for two dots, one over the other, with $R=300$ nm, $H=10$ nm, $K_u=150$ kJ/m$^3$, $x=0$, and $z=30$ nm. (a) Comparison between the analytic magnetic profile of the N\'{e}el-like skyrmion configuration and the magnetic profile obtained by the micromagnetic simulation with OOMMF. (b) Top view of the magnetization obtained by OOMMF for the  two dots. (c) Schematic representation of the front view of the magnetization of the two dots.}
\label{compareb}
\end{center}
\end{figure}

In this section we consider two disks, separated by a vertical distance  ($z>H$) and in the same axis ($x=0$). Our micromagnetic simulations show that the magnetic configuration of each dot changes from a Bloch-like skyrmion configuration to a N\'{e}el-like skyrmion configuration as the distance between the dots decreases. In a N\'{e}el-like skyrmion the  magnetization rotates in the plane parallel to the radial direction, then we propose a  magnetic profile of the form $\vec{M}_A \left(\vec{r}\right)=-M_s m_{\rho}\left(\rho\right)\hat{\rho}+M_s m_{z}\left(\rho\right)\hat{z}$ and  $\vec{M}_B \left(\vec{r}\right)=M_s m_{\rho}\left(\rho\right)\hat{\rho}+M_s m_{z}\left(\rho\right)\hat{z}$, for  the dot $A$ and for the dot $B$, respectively. We have that   $m_{\rho}^2(\rho)+m_{z}^2(\rho)=1$ and 
 $m_{z}\left(\rho\right)$ is given by Eq. (\ref{perfil2}). Figure \ref{compareb} illustrates the $z$-component of the magnetization of the magnetic profile for 
  $z=30$ nm and $K_u=150$ kJ/m$^3$.   Figure \ref{compareb}(a) illustrates a comparison between the analytic magnetic profile  and the magnetic profile obtained by the micromagnetic simulation with OOMMF. Figure \ref{compareb}(b) shows a top view of the magnetization of the two dots. Figure \ref{compareb}(c) shows a schematic representation of the front view of the magnetization of both dots.

The total magnetic energy of the system with the N\'{e}el-like skyrmion configuration for the dots, $E_{\text{tot}}^{N}$, is given by the sum
of the exchange, magnetostatic, and anisotropy
energies. The abbreviation $N$ is used  with the aim of referring to a  N\'{e}el-like skyrmion.
The exchange energy for the two dots with the $N$ configuration, $E_{\text{ex}}^N$, is the same given by the $B_2$ configuration, i.e., $E_{\text{ex}}^N=E_{\text{ex}}^{B_2}$. The self-magnetostatic
   energy in this case,  comes from the superficial and volumetric magnetic charges. Then, the self-magnetostatic
   energy is:
   \begin{ceqn}
   \begin{align}
E_{\text{m,self}}^N=&\pi \mu_0 M_{s}^{2}\ \int_{0}^{\infty}dq\left(qH+e^{-qH}-1\right)\left(\int_{0}^{R}m_{\rho}(\rho)J_{1}(q\rho)\rho d\rho\right)^2 \nonumber \\
&+\pi \mu_{0}M_{s}^{2}\int_{0}^{\infty}dq\left(1-e^{-qH}\right)\left(\int_{0}^{R}m_{z}(\rho)J_{0}(q\rho)\rho d\rho\right)^{2}
\end{align}
 \end{ceqn}
The magnetostatic interaction between the two dots is equal to:
 \begin{ceqn}
\begin{align}
E_{\text{m,int}}^N\left(x=0,z\right)=&-\pi \mu_{0}M_s \int_{0}^{\infty}dq e^{-q(H+y)}(e^{qH}-1)^2 \left(\int_{0}^{R}m_{\rho}(\rho)J_1(q\rho)\rho d\rho\right)^2 \nonumber \\
&- \pi \mu_0M_s \int_{0}^{\infty}dq e^{-q(H+y)}(e^{qH}-1)^2 \left(\int_{0}^{R}m_{z}(\rho)J_{0}(q\rho)\rho d\rho \right)^2\nonumber\\
&-2 \pi \mu_0M_s \int_{0}^{\infty}dq e^{-q(H+y)}(e^{qH}-1)^2 \int_{0}^{R}m_{z}(\rho)J_{0}(q\rho)\rho d\rho \int_{0}^{R}m_{\rho}(\rho)J_1(q\rho)\rho d\rho.
\end{align}
 \end{ceqn}
The anisotropy contribution of the $N$ bubble skyrmion configuration is equal to the $B_2$ bubble configuration, i.e., $E_{\text{ani}}^N=E_{\text{ani}}^{B_2}$.
Therefore, the total energy expression of the system with the $N$ configurations for the two dots at $x=0$ is equal to
\begin{ceqn}
\begin{equation}
E_{\text{tot}}^{N}=2E_{\text{ex}}^{N}+2E_{\text{m,self}}^{N}+2E_{\text{ani}}^{N}+E_{\text{m,int}}^{N}.
\label{skytotbc}
\end{equation} 
\end{ceqn} 
This expression, Eq. (\ref{skytotbc}), depends on the parameters $x$, $z$, $\gamma$, and $\Delta$.  Therefore, to obtain the energy
of the system, we need to minimize $E_{\text{tot}}^{N}$ as a function of the
parameters $\gamma$ and $\Delta$; for a fixed $R$, $H$, $x$, $z$, and $K_u$.

\section*{Results and Discussion}

We start with two Co/Pt magnetic dots with geometrical
parameters $R=300$ nm and $H=10$ nm. The center of the dot $A$ is set up at the origin $x=z=0$. With these
parameters, we study two scenarios in the following subsections: the
disks adjacent to each other is when the dot $B$ is at $x>2R$ and $z=0$
(dots in the same plane) and the disks vertically stacked corresponds to $x=0$ and
$z>H$ (dots in the same axis $z$). In the following sections we study  and discuss the magnetostatic interaction energy and also the dependence of the parameters of the skyrmion, the core size $b$ and the end-width size  $c$, as a function of the distance between the dots. The analytical values for the core ($b_a$) and the end-width ($c_a$) sizes were obtained by minimizing the total magnetic energies given in the last section. The numerical values for the core ($b_s$) and the end-width ($c_s$) sizes were obtained by OOMMF simulations.

\subsection{Horizontal separation between two magnetic disks with bubble skyrmion configurations,  $x>2R$ and $z=0$.}
\label{subA}

\begin{figure}[h!]
\begin{center}
\includegraphics[width=8cm]{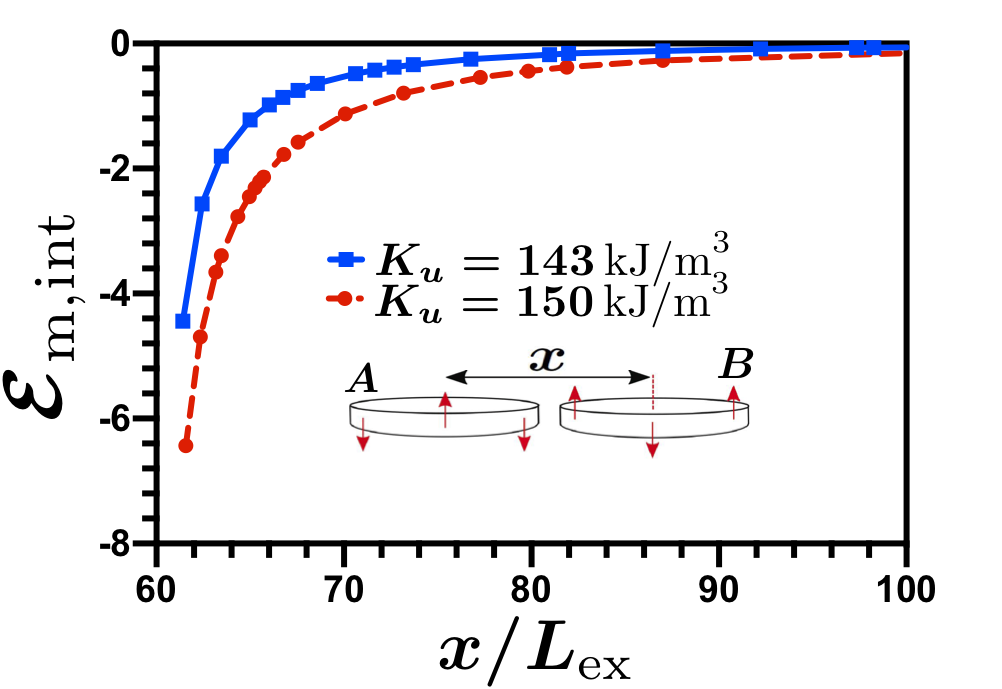}
\caption{Normalized magnetostatic interaction energy as a function of the
  horizontal separation $x$ between the two disks with the antiparallel bubble
  skyrmion configurations for different anisotropies. The symbols represent the numerical points at different anisotropies. The lines are obtained by fitting these numerical points. The square (blue solid line) 
   represents $K_u=143$ kJ/m$^3$ and the   dot (red dashed line)
 represents $K_u=150$ kJ/m$^3$. Every disk has $h=10$ nm $/L_{\text{ex}}\approx 1.02$ and $r=300$ nm$/L_{\text{ex}}\approx30.67$.}
\label{fig2}
\end{center}
\end{figure}

In this section we study the magnetostatic interaction between two
disks where the disks are one beside another, i.e., according to Fig. \ref{fig1}, $z=0$ and the distance between their centers
is $x>2R$.  Figure \ref{fig2} illustrates the
magnetostatic interaction energy normalized by $\mu_0 M_s^2
L_{\text{ex}}^3$, ${\cal{E}}_{\text{m,int}}=E_{\text{m,int}}/(\mu_0 M_s^2
L_{\text{ex}}^3)$, as a function of $x$ for $K_u=$ $143$~kJ/m$^3$ and $150$~kJ/m$^3$. In this configuration, the magnetostatic interaction energy is negative when $\sigma_A\neq \sigma_B$, so that the two skyrmions are
oriented antiparallel. Similar results were obtained with magnetic vortices in disks, where the antiparallel alignment of the vortices have the lowest magnetostatic energy \cite{Altbir2007,Porrati2005}. The orientation of the skyrmions is energetically favorable because the magnetic field
  lines produced at the edge of one of the skyrmions, can naturally and
  immediately match the orientation of the magnetization and magnetic
  field produced at the closest edge of the other skyrmion, minimizing the
  energy associated with the magnetostatic interaction (similar explanation is observed in two coupled vortices, where the magnetic cores close the magnetic field lines \cite{Altbir2007,Porrati2005}).  For this reason, we start to analyze the antiparallel configuration in this
  subsection,  as it corresponds to the configuration of lowest energy.

\begin{figure}[h!]
\begin{center}
\includegraphics[width=14cm]{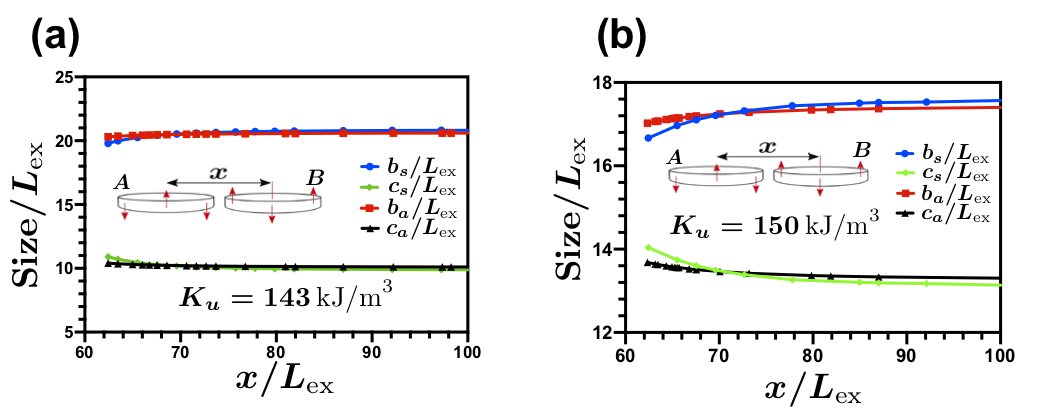}
\caption{ The core, $b$ ($b_a$ from the analytic model and $b_s$ from the micromagnetic simulations), and the edge-width, $c$ ($c_a$ from the analytic model and $c_s$ from the micromagnetic simulations), sizes of two disks with the antiparallel bubble skyrmion configurations as a function of $x$ at $h=10$ nm $/L_{\text{ex}}\approx 1.02$ and $r=300$ nm$/L_{\text{ex}}\approx30.67$. The anisotropy is $(a)$ $K_u=143$ kJ/m$^3$ and $(b)$  $K_u=150$ kJ/m$^3$.}
\label{fig3}
\end{center}
\end{figure}

To study the dependence of the core size, $b$, and the edge-width size,
$c$, of the two skyrmions as a function of $x$, we will
choose the two following regimes. The first one is when  $K_u=143$ kJ/m$^3$ and we use the configuration $B_1$ [Eq. (\ref{perfil})]. The second one is when $K_u=150$ kJ/m$^3$ and we use  the configuration $B_2$ [Eq. (\ref{perfil2})]. Figure \ref{fig3}
illustrates the parameters of the skyrmions, $b$ and $c$, as
a function of $x$. Figure \ref{fig3}(a) considers $K_u=143$ kJ/m$^3$ and Figure \ref{fig3}(b) considers  $K_u=150$ kJ/m$^3$. We observe a good agreement between the analytical calculation and the numerical simulations. When the distance
between the disks decreases, the edge-width size of the skyrmions
increases while the size of the cores decreases. This
behavior can be explained by the following argument: when the two
magnetic disks (skyrmions $A$ and $B$) are near each other,
  the magnetization directions for the edges of $A$ and $B$ are oriented
  antiparallel, allowing the magnetostatic field produce by one
  skyrmion to naturally close at the edge of the other skyrmion. This
  results in a decrease of the magnetostatic energy making the system
  to be more stable. When the skyrmions are close to each other, it is
  energetically favorable to have a relatively large edge $c$.
  However, when the disks are moved away from each other, the
  magnetostatic interaction between them is reduced compared to
  the other energy terms, so that $c$ begins to decrease until it
  reaches the value corresponding to an isolated skyrmion. In order to
  explain the behavior of the core of the skyrmion, we must first
  consider that the magnetic interaction between the core $A$ and the
  edge-width $B$ is stronger than the interaction between the core $A$ and
  the core $B$.  This is because the magnetic volume due to the magnetization perpendicular to
  the plane of the disk from the cores is less than the magnetic volume from the the edge-widths, and also because the
  distance between the core $A$ and the core $B$ is greater than the
  distance between the core $A$ and the edge-width $B$. If we
  focus on the magnetic interaction between the core of the skyrmion $A$
  and the edge-width of the skyrmion $B$ when they are close together, we see that both magnetization directions are the
  same. Such parallel configuration is not favorable, so that it is energetically
  favorable to have a small magnetic volume, i.e, the size of the
  core, $b$, should be small. For the purpose of increasing $b$, the magnetostatic interaction
should diminish, therefore, we have to increase $x$. The opposite behavior occurs for the cores of the magnetic disk that have magnetic vortices, i.e., the core radius of the vortices decreases as the distance between them increases. This discussion for the vortices is analogous to what happens with the edges of the skyrmions, which correspond to the strongest interaction in this scenario, see Refs. \cite{Altbir2007,Porrati2005}. Analogously to the edge-width size, the core size of each skyrmion
takes the value of an isolated disk when the separation distance is
large enough to consider the magnetostatic interaction energy between
the disks equal to zero.

\begin{figure}[h!]
\begin{center}
\includegraphics[width=8cm]{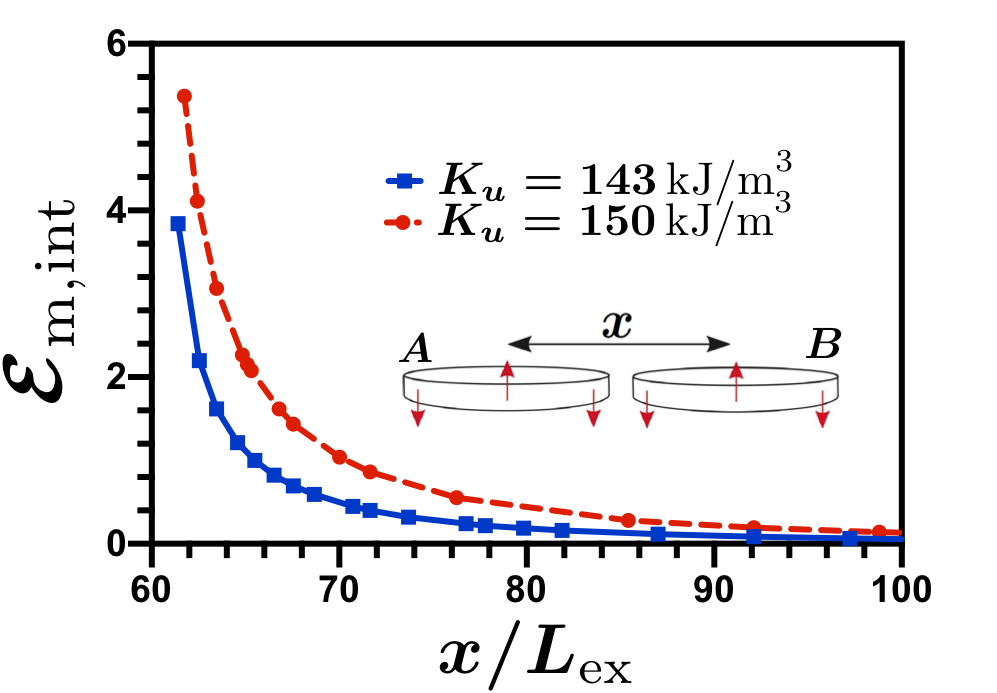}
\caption{ Normalized magnetostatic interaction energy as a function of the horizontal separation $x$ between the two disks with the parallel bubble skyrmion configurations for different anisotropies. The symbols represent the numerical points at different anisotropies. The lines are obtained by fitting these numerical points. The the square (blue solid line) represents $K_u=143$ kJ/m$^3$ and the dot (red dashed line) represents  $K_u$ =150 kJ/m$^3$. Every disk has $h=10$ nm/$L_{\text{ex}}\approx1.02$ and $r =300$ nm/$L_{\text{ex}}\approx30.67$}
\label{fig4as}
\end{center}
\end{figure}

From an application point of view, both antiparallel and parallel configurations are well worth investigating, because the core orientation of the skyrmion can be used to encode binary information and could be either up or down, depending on the information. For this reason, in addition to the previous study of the antiparallel configuration, the parallel configuration of two skyrmions with a horizontal separation is investigated. Figure \ref{fig4as} illustrates the normalized magnetostatic interaction energy, as a function of $x$ for $143$ kJ/m$^3$ and $150$ kJ/m$^3$. In this configuration, the magnetostatic interaction energy is positive.

\begin{figure}[h!]
\begin{center}
\includegraphics[width=14cm]{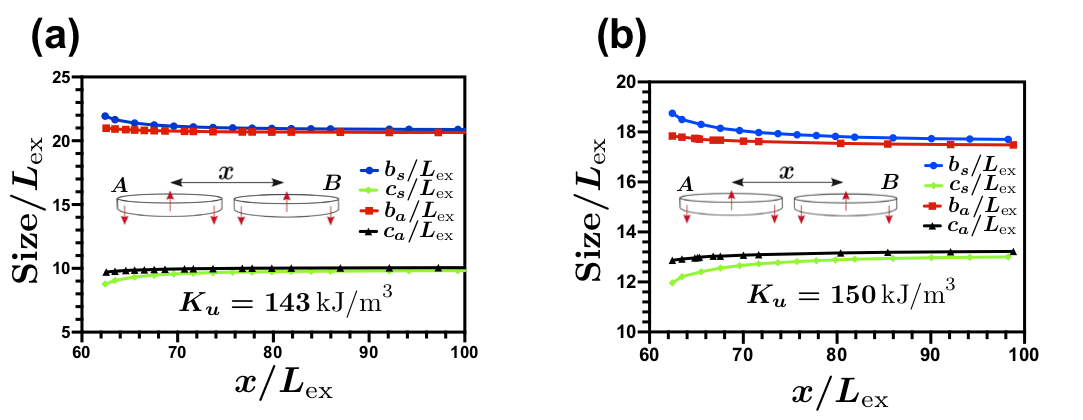}
\caption{ The core, $b$ ($b_a$ from the analytic model and $b_s$ from the micromagnetic simulations), and the edge-width, $c$ ($c_a$ from the analytic model and $c_s$ from the micromagnetic simulations), of two disks with the parallel bubble skyrmion configurations as a function of $x$ for $h =10$ nm/$L_{\text{ex}}\approx1.02$ and $r =300$ nm/$L_{\text{ex}}\approx30.67$. The anisotropy is (a)  $K_u =143$ kJ/m$^3$  and  (b)  $K_u =150$ kJ/m$^3$.}
\label{fig5as}
\end{center}
\end{figure}

Figure \ref{fig5as}  illustrates the parameters of the skyrmions, $b$  and $c$, as a function of $x$. Figure \ref{fig5as}(a) corresponds to $K_u =143$ kJ/m$^3$ and Figure \ref{fig5as}(b) corresponds to $K_u =150$ kJ/m$^3$. In both cases, when the distance between the disks decreases, the edge-width size of the skyrmions decreases while the core size increases. We observe a good agreement between the analytical calculation and the numerical simulations. This can be explained through the core-edge and edge-edge interaction. The strongest interaction is between the edges and it is unfavorable, so the parameter $c$ decreases when $x$ decreases. On the another hand, the edge-core interaction is favorable, then the parameter $b$ increases when $x$ decreases.

\subsection{Vertical separation between two magnetic disks with bubble skyrmion configurations,  $x=0$ and $z>H$}\label{subB}

\begin{figure}[h!]
\begin{center}
\includegraphics[width=14cm]{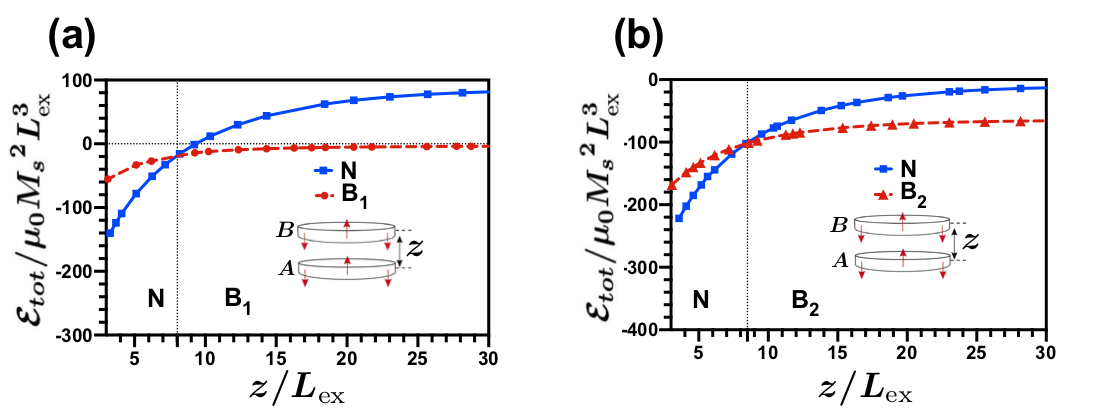}
\caption{Normalized total energy as a function of the
  vertical distance $z$ between the two disks for (a) $K_u=143$ kJ/m$^3$ and (b) $K_u=150$ kJ/m$^3$, at $x=0$. The symbols represent the different configurations: $B_1$ (dots), $B_2$ (triangles), and $N$ (squares). }
\label{x}
\end{center}
\end{figure}

In this section we study  two
disks where one disk is over the other, i.e., $x=0$ and $z>H$. To study the magnetic parameters $b$ and $c$ of the bubble skyrmions, first we need to know the magnetic configuration of the dots when the vertical distance $z$ varies. The configuration with minimum energy occurs when the bubble skyrmions are
oriented parallel to each other, since both the directions of the core
magnetization and also the directions of the edge-width magnetization
of the bubble skyrmions are the same. Analogous results have been reported in stacked ferromagnetic disks with magnetic vortices, where the parallel alignment of the vortices have the lowest magnetostatic energy\cite{Tanigaki2015,Reyes2016}. For this reason, we study the
parallel configuration, $\sigma_{A}= \sigma_{B}$, as this is the
configuration of lowest energy. Figure \ref{x}(a) shows the normalized total energy  for the two magnetic configurations ($B_1$ and $N$) as a function of $z$ for $K_u=143$ kJ/m$^3$.  We observe that the bubble skyrmion with the $B_1$ configuration is observed when the disk are isolated until the disks have a separation of $z \approx 8.02 L_{\text{ex}}$. For $z$ lower than $z=8.02 L_{\text{ex}}$, we observe the $N$ configuration for the disks. Figure \ref{x}(b) illustrates de normalized total energy for the configurations $B_2$ and $N$ as a function of $z$ for $K_u=150$ kJ/m$^3$. We observe that a distance $z= 8.41 L_{\text{ex}}$, there is a transition from the $B_2$ configuration to the $N$ configuration as $z$ decreases. Then, for both anisotropies, we observe that the bubble skyrmion configuration with a  N\'{e}el-like skyrmion configuration is stabilized by the magnetostatic interaction, without the Dzyaloshinskii-Moriya interaction, i.e., there is a transition from the  Bloch-like to the  N\'{e}el-like configuration.

    \begin{figure}[h!]
\begin{center}
\includegraphics[width=14cm]{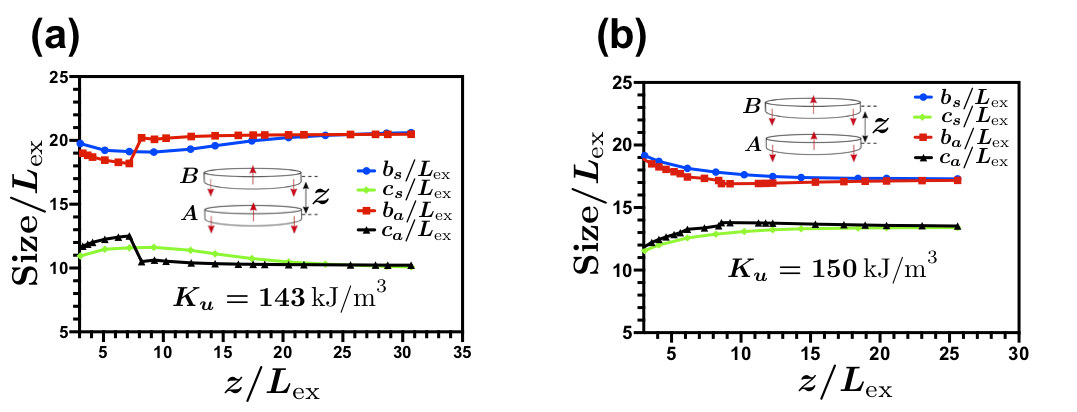}
\caption{ The core, $b$, and the edge-width, $c$, sizes of two disks with the parallel bubble skyrmion configurations as a function of $z$ at $h=10$ nm $/L_{\text{ex}}\approx 1.02$ and $r=300$ nm$/L_{\text{ex}}\approx30.67$. $(a)$ The anisotropy is $K_u=143$ kJ/m$^3$. (b)The anisotropy is $K_u=150$ kJ/m$^3$.}
\label{fig5}
\end{center}
\end{figure}

\begin{figure}[h!]
\begin{center}
\includegraphics[width=16cm]{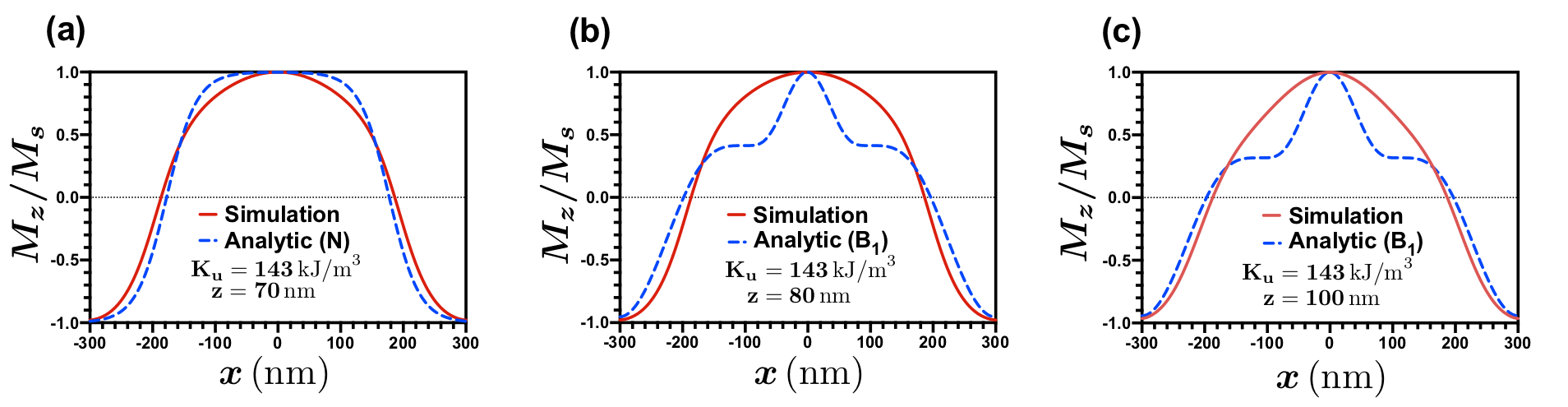}
\caption{ Component $z$ of the normalized magnetization at different vertical distance $z$ of two disks with the parallel bubble skyrmion configuration. The top row  is for $K_u=143$  kJ/m$^3$ and the bottom row is for  $K_u=150$  kJ/m$^3$.  Each disk has $h=10$ nm $/L_{\text{ex}}\approx 1.02$ and $r=300$ nm$/L_{\text{ex}}\approx30.67$.}
\label{fig5xx}
\end{center}
\end{figure}

Figure \ref{fig5}  shows the variation of the magnetic parameters
$b$ and $c$ of the two skyrmions as a function of $z$. We consider $K_u=143$ kJ/m$^3$ and $K_u=150$ kJ/m$^3$ for Fig. \ref{fig5}(a)  and Fig. \ref{fig5}(b), respectively. By decreasing the distance between the
disks, the cores sizes of the skyrmions increase for $z<7.16 L_{\text{ex}}$ for $K_a=143$ kJ/m$^3$ and all the study range of $z$ for $K_a=150$ kJ/m$^3$.
  However the edge-width sizes of the skyrmions decrease for $z<7.16 L_{\text{ex}}$ for $K_a=143$ kJ/m$^3$ and all the study range of $z$ for $K_a=150$ kJ/m$^3$.  To understand this
behavior for these zones, we observe the magnetic charges (related to the normal
  component of $\vec{M}$) on the surfaces of the magnetic disks.
We will call $B$ the upper disk and $A$ the lower disk.
Both disks have a parallel magnetic configuration.  We
consider that they are very close together so that
  $z\approx H$.  Disk $A$ has a magnetic charge $-q$ on the top
surface of the edge-width while disk $B$ has a magnetic
charge $q$ on the bottom surface of the edge-width. This configuration
is stable because the interaction between the disks reduces the
magnetic energy causing that the core sizes of the skyrmions
increase. For this reason the core sizes of the skyrmions increase
when $z$ decreases because opposite magnetic charges are attracted.
The edge-width sizes decrease because for close distance the condition $b+c=R$ occurs.  From Fig. \ref{fig5}(a) or $K_a=143$ kJ/m$^3$, we observe that the analytical model for  $z>7.16 L_{\text{ex}}$ does not reproduce the behavior of  the core and edge width sizes. The reason is that the magnetic profile for $K_a=143$ kJ/m$^3$ is more complex. Figure (\ref{fig5xx}) illustrates a comparison between the analytical profile of $m_z$ and the micromagnetic simulations for $K_u=143$  kJ/m$^3$. We observe  that the magnetic profiles of $m_z$ obtained by micromagnetic simulations, for the $B_1$ configuration, are different from the analytical profile that we consider, i.e., there is a continuous transition from the N\'{e}el configuration (with $m_{\phi}=0$ and $m_{\rho}\neq 0$) to the Bloch configuration (with $m_{\phi}\neq0$ and $m_{\rho}= 0$). 

To finalize, we did not consider the case of antiparallel configuration when one disk is over the other. The reason is that we did not observe this configuration in the micromagnetic simulation performed by OOMMF when the disks have a strongly magnetostatic interaction.

\section*{Conclusions}

In summary, by means of an analytic model and numerical
calculations, we have studied the dependence of the core and edge-width
sizes for two magnetic disks, that have a bubble skyrmion
configuration, that are interacting by the magnetostatic interaction.
By using different ansatz for the magnetic profile of a bubble skyrmion, it
was possible to obtain an expression for the magnetostatic interaction
energy between the two disks.  We observed that the magnetic
parameters that describe a skyrmion vary in different ways depending
on the location of the disks.  When the disks are
separated by a horizontal distance, the configuration with minimum energy corresponds to the skyrmions that have an
anti-parallel orientation.  Results show that if the horizontal
distance decreases, the core sizes of the skyrmions decrease and the
edge-width sizes of the skyrmions increase.  These results can be
explained by the magnetic interactions between the magnetostatic
fields created by the magnetizations of the cores and the edge-widths
of the skyrmions.  When one disk is over the other, the configuration with minimum energy corresponds to
skyrmions that have a parallel orientation. As the vertical distance
decreases, we observe that the bubble skyrmion configuration with a  N\'{e}el-like skyrmion configuration is stabilized by the magnetostatic interaction, without the Dzyaloshinskii-Moriya interaction, i.e., there is a transition from the  Bloch-like to the  N\'{e}el-like configuration. Thus, these results can be used in the fabrication
of future magnetic devices in which two or more bubble-type skyrmions
are present.

\section*{Acknowledgements}

We acknowledge financial support in Chile from FONDECYT 1161018, Financiamiento Basal para Centros Cient\'ificos y Tecnol\'{o}gicos de Excelencia FB 0807, and AFOSR Neuromorphics Inspired Science FA9550-18-1-0438. M. A. C acknowledges Conicyt-PCHA/Doctorado Nacional/2017-21171016. D. M.-A. acknowledges financial support in Chile from CONICYT, Postdoctorado FONDECYT 2018, folio 3180416.

\section*{Author contributions statement}

M. A. C., D. M-A., and S. A. carried out numerical analysis and prepared the the figures. D. M-A., J. A. V., and S. A. contributed to write the manuscript. All authors reviewed the manuscript.

\section*{Additional information}

\textbf{Competing interests:} The authors declare no competing interests.

\end{document}